\documentclass[prd,aps,showpacs,nofootinbib]{revtex4}
\usepackage{amssymb}
\usepackage{graphicx}
\usepackage{amsmath}

\newcommand{\be}{\begin{equation}}
\newcommand{\ee}{\end{equation}}
\newcommand{\bq}{\begin{eqnarray}}
\newcommand{\eq}{\end{eqnarray}}
\input{tcilatex}

\begin{document}

\title{\textbf{\ Lorentz-violating corrections on the hydrogen spectrum
induced by a non-minimal coupling}}
\author{H. Belich $^{b,e}$, T. Costa-Soares$^{c,d,e}$, M.M. Ferreira Jr.$%
^{a,e}$, J.A. Helay\"{e}l-Neto$^{c,e}$ and F. M. O. Mouchereck$^{a}$}
\affiliation{$^{a}${\small {Universidade Federal do Maranh\~{a}o (UFMA), Departamento de F%
\'{\i}sica, Campus Universit\'{a}rio do Bacanga, S\~{a}o Luiz - MA,
65085-580 - Brasil}}}
\affiliation{$^{b}${\small {Universidade Federal do Esp\'{\i}rito Santo (UFES),
Departamento de F\'{\i}sica e Qu\'{\i}mica, Av. Fernando Ferrari, S/N
Goiabeiras, Vit\'{o}ria - ES, 29060-900 - Brasil}}}
\affiliation{{\small {~}}$^{c}${\small {CBPF - Centro Brasileiro de Pesquisas F\'{\i}%
sicas, Rua Xavier Sigaud, 150, CEP 22290-180, Rio de Janeiro, RJ, Brasil}}}
\affiliation{$^{d}${\small {Universidade Federal de Juiz de Fora (UFJF), Col\'{e}gio T%
\'{e}cnico Universit\'{a}rio, av. Bernardo Mascarenhas, 1283, Bairro F\'{a}%
brica - Juiz de Fora - MG, 36080-001 - Brasil}}}
\affiliation{$^{e}${\small {Grupo de F\'{\i}sica Te\'{o}rica Jos\'{e} Leite Lopes, C.P.
91933, CEP 25685-970, Petr\'{o}polis, RJ, Brasil}}}

\begin{abstract}
The influence of a Lorentz-violating fixed background on fermions is
considered by means of a torsion-free non-minimal coupling. The
non-relativistic regime is assessed and the Lorentz-violating Hamiltonian is
determined. The effect of this Hamiltonian\ on the hydrogen spectrum is
determined to first-order evaluation (in the absence of external magnetic
field), revealing that there appear some energy shifts that modify the fine
structure of the spectrum. In the case the non-minimal coupling is
torsion-like, no first order correction shows up in the absence of an
external field; in the presence of an external field, a secondary Zeeman
effect is implied. Such effects are then used to set up stringent bounds on
the parameters of the model.
\end{abstract}

\email{belich@cbpf.br, tcsoares@cbpf.br, manojr@ufma.br, helayel@cbpf.br,
fmoucherek@yahoo.com.br}
\maketitle

\section{Introduction}

Since the pioneering work by Carroll-Field-Jackiw \cite{Jackiw},
Lorentz-violating theories have been extensively studied and used as an
effective probe to test the limits of Lorentz covariance.\textbf{\ }
Nowadays, these theories are encompassed in the framework of the Extended
Standard Model (SME), conceived by Colladay and Kosteletck\'{y} \cite%
{Colladay}\ as a possible extension of the minimal Standard Model of the
fundamental interactions. The SME admits Lorentz and CPT violation in all
sectors of interactions by incorporating tensor terms (generated possibly as
vacuum expectation values of a more fundamental theory) that account for
such a breaking. Actually, the SME model sets out as an effective model that
keeps unaffected the $SU(3)\times SU(2)\times U(1)$ gauge structure of the
underlying fundamental theory while it breaks Lorentz symmetry at the
particle frame. \ 

Concerning the gauge sector of the SME, many studies have been developed
that focus on many different respects \cite{Colladay}-\cite{Belich}. The
fermion sector has been investigated as well, initially by considering
general features (dispersion relations, plane-wave solutions, and energy
eigenvalues) \cite{Colladay}, and later by scrutinizing CTP-violating
probing experiments conceived to find out in which extent the Lorentz
violation may turn out manifest and to set up upper bounds on the breaking
parameters. The CPT theorem, valid for any local Quantum Field Theory,
predicts the equality of some quantities (life-time, mass, gyromagnetic
ratio, charge-to-mass ratio) for particle and anti-particle. Thus, in the
context of quantum electrodynamics, the most precise and sensitive tests of
Lorentz and CPT invariance refer to comparative measurement of these
quantities. A well-known example of this kind of test involves
high-precision measurement of the gyromagnetic ratio \cite{Gyro} and
cyclotron frequencies \cite{Cyclotron} for electron and positron confined in
a Penning trap for a long time. The unsuitability of the\ usual figure of
merit adopted in these works, based on the difference of the g-factor for
electron and positron, was shown in refs. \cite{Penning}, in which an
alternative figure of merit was proposed, able to constrain the
Lorentz-violating coefficients (in the electron-positron sector) to $1$\
part in $10^{20}$. Other interesting and precise experiments, also devised
to establish stringent bounds on Lorentz violation, proposed new figures of
merit involving the analysis of the hyperfine structure of the muonium
ground state \cite{Muon}, clock-comparison experiments \cite{Clock},
hyperfine spectroscopy of hydrogen and anti-hydrogen \cite{Hydrog}, and
experiments with macroscopic samples of spin-polarized matter \cite{Spin}.

The influence of Lorentz-violating and CPT-odd terms specifically on the
Dirac equation has been studied in refs. \cite{Halmi}, with the evaluation
of the nonrelativistic Hamiltonian and the associated energy-level shifts. A
similar investigation searching for deviations on the spectrum of hydrogen
has been recently performed in ref. \cite{Fernando}, where the
nonrelativistic Hamiltonian was derived directly from the modified Pauli
equation. Some interesting energy-level shifts, such as a Zeeman-like
splitting, were then reported. These results may also be used to set up
bounds on the Lorentz-violating parameters.

In another paper involving the fermion sector \cite{ACNminimo}, the
influence of a non-minimally coupled Lorentz-violating background on the
Dirac equation has been investigated. It was then shown that such a
coupling, given at the form $\epsilon _{\mu \nu \alpha \beta }\gamma ^{\mu
}v^{\nu }F^{\alpha \beta }$, is able to induce topological phases
(Aharonov-Bohm and Aharonov-Casher \cite{Casher}) at the wave function of an
electron (interacting with the gauge field and in the presence of the fixed
background). Lately, in connection with this particular effect, it has been
shown that (non-minimally coupled) particles and antiparticles develop
opposite A-Casher phases. This fact, in the context of a suitable
experiment, may be used to constrain the Lorentz-violating parameter \cite%
{Thales}. In these papers, however, it was not addressed the issue
concerning the nonrelativistic corrections induced by this kind of coupling
in an atomic system. 

The present work has as its main goal to examine the effects of the
Lorentz-violating background, whenever non-minimal coupled as in ref. \cite%
{ACNminimo}, on the Dirac equation, with special attention to its
nonrelativistic regime and possible implications on the hydrogen spectrum.
The starting point is the Dirac Lagrangian supplemented by Lorentz and
CPT-violating terms. The investigation of the nonrelativistic limit is
performed and the Lorentz-violating Hamiltonian is written down. The effect
of the background on the spectrum of hydrogen atom is then evaluated by
considering a first-order perturbation. In the absence of external magnetic
field, it is verified that three different corrections are attained, able to
modify the fine structure of the spectrum. For the case of the torsion-like
non-minimal coupling, $g_{a}\epsilon _{\mu \nu \alpha \beta }\gamma
_{5}v^{\nu }F^{\alpha \beta }$, no correction is found out. In the presence
of an external field, this term yields a Zeeman splitting proportional to
the background magnitude. The theoretical modifications here obtained are
used to set up stringent bounds on the magnitude of the corresponding
Lorentz-violating coefficient.

This paper is outlined as follows. In Sec. II, it is analyzed the influence
of the non-minimal coupling on the nonrelativistic limit of the Dirac
equation, focusing on the possible corrections induced on the spectrum of
the hydrogen atom. This is done both for a torsion-free and torsion-like
non-minimal coupling. In Sec. III, we present our Final Remarks.

\section{Non-minimal coupling to the gauge field and background}

The non-minimal coupling of the particle to\textbf{\ }the Lorentz-violating
background is here considered in two versions\textbf{:} a torsion-free and a
torsion-like coupling. We begin by analyzing the torsion-free case, which is
implemented by defining a covariant derivative with non-minimal coupling, as
below:

\begin{equation}
D_{\mu }=\partial _{\mu }+ieA_{\mu }+igv^{\nu }F_{\mu \nu }^{\ast },
\label{covader}
\end{equation}%
where $F_{\mu \nu }^{\ast }$ is the dual electromagnetic tensor $(F_{\mu \nu
}^{\ast }=\frac{1}{2}\epsilon _{\mu \nu \alpha \beta }F^{\alpha \beta }).$
In this situation, the additional term sets a non-minimal coupling of the
fermion sector to a fixed background $v^{\mu }$, responsible for the
breaking of Lorentz symmetry \cite{Jackiw} at the particle frame. The mass
dimensions of the gauge field and the coupling constant are: $\left[ A_{\mu }%
\right] =1,\left[ g\right] =-2.$ The Dirac equation with such a coupling,

\begin{equation}
(i\hbar \gamma ^{\mu }D_{\mu }-m_{e}c)\Psi =0,  \label{Dirac1}
\end{equation}%
is the starting point to investigate the influence of this background on the
dynamics of the fermionic particle. Working with the Dirac representation%
\footnote{%
In the such a representation, the Dirac matrices are written as: $\gamma
^{0}=\left( 
\begin{tabular}{ll}
$1$ & $0$ \\ 
$0$ & $-1$%
\end{tabular}%
\right) ,$ $\overset{\rightarrow }{\gamma }=\left( 
\begin{tabular}{ll}
$0$ & $\overset{\rightarrow }{\sigma }$ \\ 
$-\overset{\rightarrow }{\sigma }$ & $0$%
\end{tabular}%
\right) ,$ $\gamma _{5}=\left( 
\begin{tabular}{ll}
$0$ & $1$ \\ 
$1$ & $0$%
\end{tabular}%
\right) ,$where $\overrightarrow{\sigma }=(\sigma _{x},\sigma _{y},\sigma
_{z})$ are the Pauli matrices.} of the $\gamma $-matrices, and writing
\thinspace $\Psi $\ in terms of two-component spinors, $\Psi =\left( 
\begin{tabular}{l}
$\phi $ \\ 
$\chi $%
\end{tabular}%
\right) ,$\ there follow two coupled equations for $\phi $ and $\chi $ in
momentum space: 
\begin{eqnarray}
\left( E/c-m_{e}c-eA_{0}/c+g\overrightarrow{v}\cdot \overrightarrow{B}%
\right) \phi -\overrightarrow{\sigma }\cdot (\overrightarrow{p}-e%
\overrightarrow{A}/c+gv^{0}\overrightarrow{B}-g\overrightarrow{v}\times 
\overrightarrow{E})\chi  &=&0,  \label{phi1} \\
-\left( E/c+m_{e}c-eA_{0}/c+g\overrightarrow{v}\cdot \overrightarrow{B}%
\right) \chi +\overrightarrow{\sigma }\cdot (\overrightarrow{p}-e%
\overrightarrow{A}+gv^{0}\overrightarrow{B}-g\overrightarrow{v}\times 
\overrightarrow{E})\phi  &=&0.  \label{phi2}
\end{eqnarray}%
To investigate the low-energy behavior of this system, the natural option is
to search for its nonrelativistic limit, where the energy is given as $%
E=m_{e}c^{2}+H,$ with $H$ being the nonrelativistic Hamiltonian. Writing the
weak component ($\chi )$ in terms of the strong one ($\phi )$, the following
equation for $\phi $ holds: $\left( H/c-eA_{0}/c+g\overrightarrow{v}\cdot 
\overrightarrow{B}\right) \phi =\frac{1}{2m_{e}c}\left( \overrightarrow{%
\sigma }\cdot \overrightarrow{\Pi }\right) \left( \overrightarrow{\sigma }%
\cdot \overrightarrow{\Pi }\right) \phi ,$ where the generalized canonical
moment is defined as $\overrightarrow{\Pi }=\left( \overrightarrow{p}-e%
\overrightarrow{A}/c+gv^{0}\overrightarrow{B}-g\overrightarrow{v}\times 
\overrightarrow{E}\right) $. After some algebra, the nonrelativistic
Hamiltonian for the particle comes out: 
\begin{eqnarray}
H &=&\biggl\{\left[ \frac{1}{2m_{e}}(\overrightarrow{p}-\frac{e}{c}%
\overrightarrow{A})^{2}+eA_{0}-\frac{e\hbar }{2m_{e}c}(\overrightarrow{%
\sigma }\cdot \overrightarrow{B})\right] +\frac{g^{2}}{2m_{e}}(%
\overrightarrow{v}\times \overrightarrow{E})^{2}+\frac{\hbar }{2m_{e}}gv_{0}%
\overrightarrow{\sigma }\cdot (\overrightarrow{\nabla }\times 
\overrightarrow{B})  \notag \\
&&-\frac{g\hbar }{2m_{e}}\overrightarrow{\sigma }\cdot \lbrack 
\overrightarrow{\nabla }\times (\overrightarrow{v}\times \overrightarrow{E}%
)]+\frac{gv_{0}}{m_{e}}(\overrightarrow{p}-\frac{e}{c}\overrightarrow{A}%
)\cdot \overrightarrow{B}-\frac{g}{m_{e}}(\overrightarrow{p}-\frac{e}{c}%
\overrightarrow{A})\cdot (\overrightarrow{v}\times \overrightarrow{E})-\frac{%
g^{2}v_{0}}{m_{e}}\overrightarrow{B}\cdot (\overrightarrow{v}\times 
\overrightarrow{E})\biggr\},  \label{H1}
\end{eqnarray}%
In the expression above, there appears the Pauli Hamiltonian (between
brackets) corrected by the terms that compose the Lorentz-violating
Hamiltonian, $H_{LV\text{ }},$ which truly constitutes our object of
interest. The purpose is to investigate the contribution of the Hamiltonian
with\ Lorentz-violation $(H_{LV})$ in the states of the hydrogen atom. Such
a calculation will be initially performed for the case of a free hydrogen
atom (without external field, $\overrightarrow{A}=0$), for which only three
terms contribute. For all the terms that do not involve the spin operator,
we shall use the hydrogen 1-particle wave functions $\left( \Psi \right) $
labeled in terms of the quantum numbers $n,l,m$, $\Psi _{nlm}(r,\theta ,\phi
)=R_{nl}(r)\Theta _{lm}(\theta )\Phi _{m}(\phi ),$ whereas the evaluation of
the terms involving $\overrightarrow{\sigma }$\ requires the use of the wave
function $\Psi _{nljm_{j}m_{s}},$ with $j,m_{j\text{ }}$being the quantum
numbers suitable to deal with the addition of angular momentum. Here, $%
r,\theta ,\phi $ are spherical coordinates.

As our initial evaluation\footnote{%
It is worthwhile to mention here that all calculations have been carried out
in the Gaussian unit system, adopted through this work.}, we consider the
first-order correction induced by the term $g^{2}(\overrightarrow{v}\times 
\overrightarrow{E})^{2}/2m,$ namely:

\begin{equation}
\Delta E_{1}=\frac{g^{2}}{2m_{e}}\int \Psi _{nlm}^{\ast }(\overrightarrow{v}%
\times \overrightarrow{E})^{2}\Psi _{nlm}d^{3}r.  \label{Energ}
\end{equation}%
To solve it, we write $(\overrightarrow{v}\times \overrightarrow{E}%
)^{2}=v^{2}E^{2}-(\overrightarrow{v}\cdot \overrightarrow{E})^{2}$ and take
the Coulombian electric field given by $\overrightarrow{E}=-e\widehat{r}%
/r^{2},$ so that the result is: 
\begin{equation}
\Delta E_{1}=\frac{g^{2}e^{2}}{2m_{e}}[v^{2}\left\langle
nlm|1/r^{4}|nlm\right\rangle -\left\langle nlm|(\overrightarrow{v}\cdot 
\widehat{r})^{2}/r^{4}|nlm\right\rangle ].
\end{equation}

In spherical coordinates, $\overrightarrow{v}\cdot \widehat{r}=v_{x}\sin
\theta \cos \phi +v_{y}\sin \theta \sin \phi +v_{z}\cos \theta ,$ which
leads us to:

\begin{equation*}
\Delta E_{1}=\frac{g^{2}e^{2}}{4m_{e}}\left[ \overline{\left( \frac{1}{r^{4}}%
\right) }(v_{x}^{2}+v_{y}^{2}+2v_{z}^{2})+(v_{x}^{2}+v_{y}^{2}-2v_{z}^{2})%
\left\langle nlm|\frac{\cos ^{2}\theta }{r^{4}}|nlm\right\rangle \right] .
\end{equation*}%
Considering the intermediate result, 
\begin{equation*}
\left\langle nlm|\frac{\cos ^{2}\theta }{r^{4}}|nlm\right\rangle =\overline{%
\left( \frac{1}{r^{4}}\right) }\left[ \frac{(l^{2}-m^{2})}{(2l-1)(2l+1)}+%
\frac{(l^{2}-m^{2}+2l+1)}{\left( 2l+3\right) (2l+1)}\right] ,
\end{equation*}%
the following energy correction is obtained for the case the background is
aligned along the z-axis$\left( \overrightarrow{v}=v_{z}\overset{\symbol{94}}%
{z}\right) $:

\begin{equation}
\Delta E_{1}=\frac{g^{2}e^{2}v_{z}^{2}}{4m_{e}}\overline{\left( \frac{1}{%
r^{4}}\right) }\left[ 1-\left( \frac{(l^{2}-m^{2})}{(2l-1)(2l+1)}+\frac{%
(l^{2}-m^{2}+2l+1)}{\left( 2l+3\right) (2l+1)}\right) \right] .  \label{CE1}
\end{equation}%
where $\overline{\left( 1/r^{4}\right) }=\left\langle
nlm|1/r^{4}|nlm\right\rangle
=3[1-l(l+1)/3n^{2}]/[n^{3}a_{0}^{4}(l+3/2)(l+1)(l+1/2)l(l-1/2)]$ is a
well-known result for the hydrogen system. Here, $a_{0}=\hbar
^{2}/e^{2}m_{e} $ is the Bohr radius ($a_{0}=0.0529nm$). This result shows
that the non-minimal coupling is able to remove the accidental degeneracy,
regardless the spin-orbit interaction. This effect, therefore, implies a
modification on the fine structure of the spectrum. The order of magnitude
of this correction is given by the ratio $%
g^{2}v_{z}^{2}e^{2}/(m_{e}a_{0}^{4}),$ which\ is numerically $2\times
10^{53}\left( gv_{z}\right) ^{2}eV.$ Considering that spectroscopic
experiments \ are able to detect effects of one part in $10^{10}$ in the
spectrum$,$ the correction (\ref{CE1}) may not be larger than $10^{-10}eV,$
which implies an upper bound for the product $gv_{z}$, namely:\ $gv_{z}\leq
10^{-32}.$

In the absence of an external magnetic field, the next term to be taken into
account is $\frac{g}{m_{e}}(\overrightarrow{p}-e\overrightarrow{A})\cdot (%
\overrightarrow{v}\times \overrightarrow{E}),$ whose non-trivial part is $%
\frac{g}{m_{e}}\overrightarrow{p}\cdot (\overrightarrow{v}\times 
\overrightarrow{E}).$ Hence, the first-order energy correction is: 
\begin{equation}
\Delta E_{2}=-i\hbar m_{e}\int \Psi _{nlm}^{\ast }\nabla \cdot (%
\overrightarrow{v}\times \overrightarrow{E})\Psi _{nlm}d^{3}r=-i\hbar \frac{g%
}{m_{e}}\int \Psi ^{\ast }[\nabla \cdot (\overrightarrow{v}\times 
\overrightarrow{E})]\Psi d^{3}r-i\hbar \frac{g}{m_{e}}\int \Psi ^{\ast }(%
\overrightarrow{v}\times \overrightarrow{E})\cdot \nabla \Psi d^{3}r.
\label{E2}
\end{equation}%
Taking the gradient of $\Psi $ in spherical coordinates, and the scalar
product with $(\overrightarrow{v}\times \overrightarrow{E}),$ many terms are
obtained\ that depend linearly on $\sin \phi $, $\cos \phi $ or $\sin 2\phi ,
$ except for two of them. These are the ones that survive after the angular
integration is performed. The remaining expression is: 
\begin{equation}
\Delta E_{2}=\frac{egv_{z}m\hbar }{m_{e}}\int R_{nl}^{\ast }(r)\Theta
_{lm}^{\ast }(\theta )\frac{1}{r^{3}}R_{nl}(r)\Theta _{lm}(\theta
)r^{2}dr\sin \theta d\theta =\frac{egv_{z}m\hbar }{m_{e}}\overline{\left( 
\frac{1}{r^{3}}\right) }.
\end{equation}%
which can be explicitly written as: 
\begin{equation}
\Delta E_{2}=\frac{egv_{z}\hbar }{m_{e}}\frac{m}{a_{0}^{3}n^{3}l(l+1/2)(l+1)}%
,  \label{CE2}
\end{equation}%
where the well-known result $\overline{\left( 1/r^{3}\right) }%
=[a_{0}^{3}n^{3}l(l+1/2)(l+1)]^{-1}$ has been used$.$ A previous superficial
examination of eq. (\ref{E2}) could lead to the misleading expectation of a
vanishing result, once it consists of the average of a linear function of
the momentum $\left( p\right) $ on the state $\Psi $. Yet, in the
development of this expression, there arise the angular momentum, $%
\overrightarrow{L}=\overrightarrow{r}\times \overrightarrow{p},$ whose
expectation value in a bound state is generally non-vanishing, justifying
the result of eq. (\ref{CE2}). The order of magnitude of this correction is $%
e\hbar gv_{z}/(m_{e}a_{0}^{3}),$ whose numerical value is $2\times
10^{27}\left( gv_{z}\right) eV.$ Taking into account the possibility of \
detection of one part in $10^{10},$ we arrive at the following bound for the
parameters:\ $gv_{z}\leq 10^{-19}.$

In order to evaluate the correction associated with the terms involving the
spin operator, it is necessary to work with the wave functions $\Psi
_{nljm_{j}m_{s}}=\psi _{nljm_{j}}(r,\theta ,\phi )\chi _{sm_{s}},$ suitable
to treat the situations where there occurs addition of angular momenta ($%
J=L+ $ $S$), with $n,l,j,m_{j}$ being the associated quantum numbers.
Considering the free hydrogen atom, the first non-null spin term is $%
\overrightarrow{\sigma }\cdot \lbrack \overrightarrow{\nabla }\times (%
\overrightarrow{v}\times \overrightarrow{E})],$ which implies the following
first-order correction:\ 

\begin{equation}
\Delta E_{3}=\frac{g\hbar }{2m_{e}}\langle nljm_{j}m_{s}|\overrightarrow{%
\sigma }\cdot (\overrightarrow{\nabla }\times (\overrightarrow{v}\times 
\overrightarrow{E}))|nljm_{j}m_{s}\rangle .  \label{EB}
\end{equation}%
For the case of the Coulombian electric field, $\overrightarrow{\sigma }%
\cdot \lbrack \overrightarrow{\nabla }\times (\overrightarrow{v}\times 
\overrightarrow{E})]=2e(\overrightarrow{\sigma }\cdot \overrightarrow{v}%
)/r^{3}-e(\overrightarrow{v}\cdot \overrightarrow{\nabla })(\overrightarrow{%
\sigma }\cdot \overrightarrow{r})/r^{3}.$ After some algebraic
manipulations, one obtains:

\begin{equation}
\Delta E_{3}=\frac{ge\hbar }{m_{e}}\langle nljm_{j}m_{s}|(\overrightarrow{%
\sigma }{}\cdot \overrightarrow{v})/r^{3}-(\overrightarrow{f}\cdot 
\overrightarrow{\sigma })|nljm_{j}m_{s}\rangle ,  \label{EB2}
\end{equation}%
with: $f_{x}=e(-v_{x}+3v_{x}\sin ^{2}\theta \cos ^{2}\phi +3v_{y}\sin
^{2}\theta \cos \phi \sin \phi +v_{z}\cos \theta \sin \theta \cos \phi
)/r^{3};$ $f_{y}=e(-v_{y}+3v_{y}\sin ^{2}\theta \sin ^{2}\phi +3v_{x}\sin
^{2}\theta \cos \phi \sin \phi +v_{z}\cos \theta \sin \theta \cos \phi
)/r^{3};$ $f_{z}=e(-v_{z}+3v_{z}\cos ^{2}\theta +3v_{x}\sin \theta \cos
\theta \cos \phi +v_{y}\cos \theta \sin \theta \cos \phi )/r^{3}.$

\bigskip To complete this calculation, it is necessary to write the $%
|jm_{j}\rangle $ kets in terms of the spin eigenstates $|mm_{s}\rangle ,$
which is done by means of the general expression: $|jm_{j}\rangle
=\dsum\limits_{m,m_{s}}\langle mm_{s}|jm_{j}\rangle $ $|mm_{s}\rangle ,$
where $\langle mm_{s}|jm_{j}\rangle $ are the Clebsch-Gordon coefficients.
Evaluating such coefficients for the case $j=l+1/2,m_{j}=m+1/2,$ one has: $\
|jm_{j}\rangle =\alpha _{1}|m\uparrow \rangle +\alpha _{2}|m+1\downarrow
\rangle ;$ one the other hand, for $j=l-1/2,m_{j}=m+1/2,$ one obtains: $%
|jm_{j}\rangle =\alpha _{2}|m\uparrow \rangle -\alpha _{1}|m+1\downarrow
\rangle ,$ with: $\alpha _{1}=\sqrt{(l+m+1)/(2l+1)},\alpha _{2}=\sqrt{%
(l-m)/(2l+1)}.$ Taking now into account the orthonormalization relation $%
\langle m^{\prime }m_{s}^{\prime }|mm_{s}\rangle =\delta _{m^{\prime
}m}\delta _{m_{s}^{\prime }m_{s}},$ it is possible to show that eq.\ (\ref%
{EB}) leads to:

\begin{equation}
\Delta E_{3}=\pm \frac{3e\hbar gv_{z}}{2m_{e}}\frac{m_{j}}{%
a_{0}^{3}n^{3}l(l+1/2)(l+1)\left( 2l+1\right) }\left\{ 1-\left( \frac{%
(l^{2}-m^{2})}{(2l-1)(2l+1)}+\frac{(l^{2}-m^{2}+2l+1)}{\left( 2l+3\right)
(2l+1)}\right) \right\} ,
\end{equation}%
where the positive and negative signs correspond to $j=l+1/2$ and $j=l-1/2,$
respectively; it was also used: $\langle nljm_{j}m_{s}|\sigma
_{z}|nljm_{j}m_{s}\rangle =\pm m_{j}/(2l+1)$, $\langle nljm_{j}m_{s}|\sigma
_{x}|nljm_{j}m_{s}\rangle =\langle nljm_{j}m_{s}|\sigma
_{y}|nljm_{j}m_{s}\rangle =0,$ and the expression for $\overline{\left(
1/r^{3}\right) }.$ The order of magnitude of this correction is $%
gv_{z}e\hbar /(m_{e}a_{0}^{3}),$ the same of the correction $\Delta E_{2}.$

Next, we still consider an external fixed field and we evaluate the
corrections induced by it. In principle, three terms of the Hamiltonian (\ref%
{H1}) might yield non-zero contributions in the presence of a magnetic
field, namely: $\Delta E_{1B}=\frac{gv_{0}}{m_{e}}\left\langle nlm|(%
\overrightarrow{p}-e\overrightarrow{A})\cdot \overrightarrow{B}%
|nlm\right\rangle ,$ $\Delta E_{2B}=-\frac{eg}{m_{e}c}\left\langle nlm|%
\overrightarrow{A}\cdot (\overrightarrow{v}\times \overrightarrow{E}%
)|nlm\right\rangle ,$ $\Delta E_{3B}=-\frac{g^{2}v_{0}}{m_{e}}\left\langle
nlm|\overrightarrow{B}\cdot (\overrightarrow{v}\times \overrightarrow{E}%
)|nlm\right\rangle .$ For a fixed magnetic field along the z-axis, $%
\overrightarrow{B}=B_{0}\widehat{z}$, the vector potential in the symmetric
gauge reads: $\overrightarrow{A}=-B_{0}(y/2,-x/2,0).$ Concerning the first
term, only the product $\overrightarrow{A}\cdot \overrightarrow{B}$ could
provide a non-trivial contribution, once the evaluation of the product $%
\overrightarrow{p}\cdot \overrightarrow{B}$ on the wave function obviously\
vanish. After a simple inspection, one gets $\Delta E_{1B}=\frac{gv_{0}}{%
m_{e}}\left\langle nlm|\overrightarrow{A}\cdot \overrightarrow{B}%
|nlm\right\rangle =0.$

In order to solve the second term, we should write $(\overrightarrow{v}%
\times \overrightarrow{E})=$ $-\frac{e}{r^{2}}[(v_{y}\cos \theta -v_{z}\sin
\theta \sin \phi )\widehat{i}+(v_{z}\sin \theta \cos \phi -v_{x}\cos \theta )%
\widehat{j}+(v_{z}\sin \theta \sin \phi -v_{y}\sin \theta \cos \phi )%
\widehat{k}.$ The explicit calculation of this term\ yields a trivial
result. Finally, it remains to evaluate the third term, which turns out to
be also vanishing. We thus verify that the magnetic field does not yield any
correction associated with the background; it only leads to the well-known
Zeeman effect. This is the situation for the torsion free coupling.

Another possible way to couple the Lorentz-violating background ($v^{\mu }$)
to the fermion field is by proposing a torsion-like non-minimal coupling, 
\begin{equation}
D_{\mu }=\partial _{\mu }+eA_{\mu }+ig_{a}\gamma _{5}v^{\nu }F_{\mu \nu
}^{\ast },
\end{equation}%
which has a chiral character, and has been examined in ref. \cite{ACNminimo}
as well.

Writing the spinor $\Psi $\ in terms of the so-called small and large
components in much the same way as it was done in the previous case, there
follow two coupled equations for the 2-component spinors $\phi ,\chi ,$ 
\begin{eqnarray}
\left[ \left( E/c-mc-eA_{0}/c\right) -g_{a}\overrightarrow{\sigma }\cdot
\left( v^{0}\overrightarrow{B}-\overrightarrow{v}\times \overrightarrow{E}%
\right) \right] -\phi \lbrack \overrightarrow{\sigma }\cdot \left( 
\overrightarrow{p}-e\overrightarrow{A}/c\right) -g_{a}\overrightarrow{v}%
\cdot \overrightarrow{B}]\chi =0, &&  \label{phi4} \\
\lbrack \overrightarrow{\sigma }\cdot \left( \overrightarrow{p}-e%
\overrightarrow{A}/c\right) +g_{a}\overrightarrow{v}\cdot \overrightarrow{B}%
]\phi -\left[ \left( E/c+mc-eA_{0}/c\right) -g_{a}\overrightarrow{\sigma }%
\cdot \left( v^{0}\overrightarrow{B}-g_{a}\overrightarrow{v}\times 
\overrightarrow{E}\right) \right] \chi &=&0,
\end{eqnarray}%
from which we can read the weak component in terms of the strong one, $\chi =%
\frac{1}{2m_{e}}\left[ \overrightarrow{\sigma }\cdot \left( \overrightarrow{p%
}-\frac{e}{c}\overrightarrow{A}\right) +g_{a}\overrightarrow{v}\cdot 
\overrightarrow{B}\right] \phi .$ It is then possible to write the Pauli
equation, 
\begin{equation}
\left( H/c-eA_{0}/c-g_{a}\overrightarrow{\sigma }\cdot \left( v^{0}%
\overrightarrow{B}-\overrightarrow{v}\times \overrightarrow{E}\right)
\right) \phi =\left[ \overrightarrow{\sigma }\cdot \left( \overrightarrow{p}-%
\frac{e}{c}\overrightarrow{A}\right) -g_{a}\overrightarrow{v}\cdot 
\overrightarrow{B}\right] \frac{1}{2m_{e}}\left[ \overrightarrow{\sigma }%
\cdot \left( \overrightarrow{p}-\frac{e}{c}\overrightarrow{A}\right) +g_{a}%
\overrightarrow{v}\cdot \overrightarrow{B}\right] \phi ,
\end{equation}%
whose structure reveals as canonical generalized moment the usual relation,$%
\overrightarrow{\text{ }\Pi }=(\overrightarrow{p}-\frac{e}{c}\overrightarrow{%
A}).$ Simplifying the equation above, the nonrelativistic Hamiltonian takes
the form: 
\begin{equation}
H=\left[ \frac{(\overrightarrow{p}-e\overrightarrow{A}/c)^{2}}{2m_{e}}%
+eA_{0}-\frac{e\hbar }{2m_{e}c}(\overrightarrow{\sigma }\cdot 
\overrightarrow{B})\right] +g_{a}v_{0}c\overrightarrow{\sigma }\cdot 
\overrightarrow{B}-g_{a}c\overrightarrow{\sigma }\cdot (\overrightarrow{v}%
\times \overrightarrow{E})-\frac{g_{a}}{2m_{e}}(\overrightarrow{v}\cdot 
\overrightarrow{B})^{2}.  \label{H2}
\end{equation}%
This Hamiltonian has yet two additional terms, $(\overrightarrow{\sigma }%
\cdot \overrightarrow{p})(\overrightarrow{v}\cdot \overrightarrow{B})-(%
\overrightarrow{v}\cdot \overrightarrow{B})(\overrightarrow{\sigma }\cdot 
\overrightarrow{p}),$ which are equal (canceling each other) for the case of
a uniform magnetic field. They will not be considered here.

In the absence of magnetic field, only the term $\overrightarrow{\sigma }%
\cdot (\overrightarrow{v}\times \overrightarrow{E})$ contributes for the
energy, implying the following correction:

\begin{equation}
\Delta E_{\sigma }=g_{a}\langle nljm_{j}m_{s}|\overrightarrow{\sigma }\cdot (%
\overrightarrow{v}\times \overrightarrow{E})|nljm_{j}m_{s}\rangle .
\end{equation}%
Considering that $\overrightarrow{\sigma }\cdot (\overrightarrow{v}\times 
\overrightarrow{E})=$ $-\frac{e}{r^{2}}[(v_{y}\cos \theta -v_{z}\sin \theta
\sin \phi )\sigma _{x}+(v_{z}\sin \theta \cos \phi -v_{x}\cos \theta )\sigma
_{y}+(v_{z}\sin \theta \sin \phi -v_{y}\sin \theta \cos \phi )\sigma _{z},$
and the action of the spin operators on the kets $|nljm_{j}m_{s}\rangle ,$
it is easy to note that: $\Delta E_{\sigma }=0.$ Hence, the non-minimal
pseudoscalar coupling yields no background contribution for the energy
levels.

Now, the presence of an external magnetic field shall be taken into account.
In this case, there appears a non-zero new contribution associated with the
term $cg_{a}v_{0}\overrightarrow{\sigma }\cdot \overrightarrow{B},$ which
generates a Zeeman splitting of the levels, whose separation is linear on
the product $cg_{a}v_{0}.$ For the case the magnetic field is aligned with
the z-axis, the implied energy correction is $\Delta
E_{1B}=cg_{a}v_{0}B_{0}\langle nljm_{j}m_{s}|\sigma
_{z}|nljm_{j}m_{s}\rangle ,$ which yields:

\begin{equation}
\Delta E_{1B}=\pm g_{a}v_{0}cB_{0}\frac{m_{j}}{2l+1},
\end{equation}%
where the positive and negative signs correspond to $j=l+1/2$ and $j=l-1/2,$
respectively. \ This is exactly the same pattern of splitting of the Zeeman
effect, here with amplitude given as $g_{a}v_{0}B_{0}.$\ Hence, besides the
usual Zeeman effect, there occurs this secondary Zeeman splitting that
implies a correction to the effective splitting. The last term of eq. (\ref%
{H2}) only implies a constant correction on all levels, which does not lead
to any change in the spectrum. The magnitude of this correction is
proportional to $g_{a}v_{0}cB_{0}.$ If such an effect is not detectable for
a magnetic strength of $1$ $G$, it should not imply a correction larger than 
$10^{-10}eV,$ so that the bound $g_{a}v_{0}\leq 10^{-18}$ is attained.

\section{Final Remarks}

In this work, we have studied low-energy effects of a Lorentz-violating
background (non-minimally coupled to the fermion and gauge fields) on a
nonrelativistic system. Indeed, the nonrelativistic limit has been worked
out and the Lorentz-violating Hamiltonian (derived from the non-minimal
coupling) evaluated. The first-order corrections induced on the energy
levels of the hydrogen atom have been determined. As a result, we have
observed effective shifts on the hydrogen spectrum, both in the presence and
absence of an external magnetic field. In the absence of the external
magnetic field, the term $\epsilon _{\mu \nu \alpha \beta }\gamma ^{\mu
}v^{\nu }F^{\alpha \beta }$ induces three different corrections, all of them
implying modifications on the fine structure of the spectrum. \ This result
indicates the breakdown of the accidental degeneracy, with the energy
depending on $l,m$ quantum numbers. Stipulating $10^{-10}eV$ \ as the
magnitude of a maximally undetectable change in the spectrum, we have set up
an upper bound on the product of parameters: $gv_{z}\leq 10^{-32}.$

In the case of the torsion-like non-minimal coupling, no correction is
implied in the absence of\ external magnetic field; on the other hand, in
the presence of such a fixed field, a secondary Zeeman effect is obtained.
Considering that such a correction should be smaller than $10^{-10}eV,$ an
upper bound is set up for the product, namely: $g_{a}v_{0}\leq 10^{-18}$.
These results show that Lorentz violation in the context of the non-minimal
coupling  regarded here turns out as a real negligible effect. 

\begin{acknowledgments}
MMFJr and JAHN express their gratitude to CNPq (Conselho Nacional de
Desenvolvimento Cient\'{\i}fico e Tecnol\'{o}gico) and FMOM acknowledges
FAPEMA\ (Funda\c{c}\~{a}o de Amparo \`{a} Pesquisa do Estado do Maranh\~{a}%
o) for financial support.
\end{acknowledgments}

\end{document}